\documentclass[prl,amsmath,twocolumn,superscriptaddress,preprintnumbers]{revtex4-1}
\usepackage{graphicx}
\usepackage{xcolor,soul}

\begin{document}

\preprint{TTP23-022}

\title{
Falsifying Pati-Salam models with LIGO
}

\author{Peter Athron}
\email{peter.athron@njnu.edu.cn}
\affiliation{Department  of  Physics  and  Institute  of  Theoretical  Physics,  Nanjing  Normal  University, Nanjing, Jiangsu 210023, China}

\author{Csaba Bal\'azs}
\email{csaba.balazs@monash.edu}
\affiliation{School of Physics and Astronomy, Monash University, Melbourne 3800 Victoria, Australia}

\author{Tom\'as E. Gonzalo}
\email{tomas.gonzalo@kit.edu}
\affiliation{Institute for Theoretical Particle Physics (TTP), Karlsruhe Institute of Technology (KIT), 76128 Karlsruhe, Germany}

\author{Matthew Pearce}
\email{matthew.pearce1@monash.edu}
\affiliation{School of Physics and Astronomy, Monash University, Melbourne 3800 Victoria, Australia}

\begin{abstract}
We demonstrate that existing gravitational wave data from LIGO already places constraints on well motivated Pati-Salam models that allow the Standard Model to be embedded within grand unified theories. For the first time in these models we also constrain the parameter space by requiring that the phase transition completes, with the resulting constraint being competitive with the limits from LIGO data.  Both constraints are complementary to the LHC constraints and can exclude scenarios that are much heavier than can be probed in colliders.  Finally we show that results from future LIGO runs, and the planned Einstein telescope, will substantially increase the limits we place on the parameter space.

\end{abstract}

\maketitle

The observation of Gravitational Waves (GWs) by the LIGO observatory has opened a new window to explore the dynamics of the early Universe~\cite{LIGOScientific:2016aoc, LIGOScientific:2017vwq, LIGOScientific:2017ync, LIGOScientific:2018mvr}.  This is because GWs produced in the early universe may be detectable by current terrestrial interferometers such as LIGO/VIRGO~\cite{Shoemaker:2019bqt}, KAGRA~\cite{KAGRA:2018plz} or pulsar timing arrays~\cite{NANOGrav:2023hvm,NANOGrav:2023gor,Antoniadis:2023rey,Antoniadis:2023zhi}, or future ground-based or satellite missions, such as LISA~\cite{LISA:2017pwj}, the Einstein Telescope~\cite{Maggiore:2019uih, Punturo:2010zz} and others~\cite{Badurina:2019hst, Reitze:2019iox, Evans:2021gyd}.  The observation, or lack thereof, of such GWs will contribute to the exploration of beyond the Standard Model (BSM) theories, by probing high energies so far unreachable by other experiments, such as colliders.  

GWs produced by the collision of bubbles in phase transitions (PTs)~\cite{Kosowsky:1991ua, Kosowsky:1992rz} have been the focus of much recent research by the community (see e.g. Ref.~\cite{Athron:2023xlk} and references within), such as new physics contributions to the electroweak (EW) phase transition~\cite{Beniwal:2017eik, Kurup:2017dzf, Ellis:2022lft}. These transitions, however, do not typically produce GWs in the right frequency range for LIGO/VIRGO.  Rather they can be explored by future space based missions, such as LISA. Grand Unified Theories (GUTs) are amongst the best motivated and most notable BSM theories that undergo phase transitions capable of producing GWs. GUTs predict a plethora of new states and phase transitions in the intermediate energies between the EW scale and some unification scale $M_{\rm GUT} \sim 10^{16}$ GeV, which produce a rich phenomenology for searches at particle physics experiments, astrophysical, and cosmological observatories~\cite{Deppisch:2017xhv,Croon:2019kpe}. As the resulting GW frequency scales with the typical energy of the PT, these intermediate scales in GUTs can predict GWs at a wide range of frequencies. Therefore, GWs can be used to probe the high-energetic PTs predicted by GUTs and thus impose strong constraints on various models of unification.

The candidate GUT with the best properties to produce visible GWs at LIGO/VIRGO is the family of Pati-Salam (PS) models~\cite{Pati:1974yy}. These PS models derive from a fully unified model at high energies (SO(10), E6, etc), but can exist at a lower energy compatible with the frequency range of LIGO, as they can easily avoid contributions to the decay of nucleons.  GW spectra from GUT PTs have been studied in PS models~\cite{Croon:2018kqn, Huang:2020bbe}, and the low-energy child of PS, the left-right symmetric model (LRSM)~\cite{Li:2020eun,Brdar:2019fur}.  These studies focused on the visibility of the predicted GW spectra at future observatories, such as the Einstein Telescope or the Cosmic Explorer. To the best of our knowledge, no previous study of PS models (or any GUT-inspired model for that matter) has focused on the predictions for existing programs such as LIGO/VIRGO, which we do here~\footnote{However, some unconstrained dark sector models can predict GW signals in the frequency of LIGO/VIRGO~\cite{Huang:2021rrk}.}. In this work we demonstrate this using a rigorous analysis of the phase transition in a PS model. This includes using two-loop renormalisation group equations (RGEs) and one-loop thresholds to extract model parameters. Furthermore, we account for recently identified effects from the finite duration of the gravitational wave source. Lastly we carefully handle the strongly supercooled phase transitions that lead to the strongest signals in our results.

We show that LIGO/VIRGO data can already place constraints on models within the PS family. We explicitly demonstrate this by constructing a specific example that gives rise to strong first order phase transitions and placing limits from the LIGO/VIRGO data on the parameter space of this model. We additionally show, for the first time in this class of models, that imposing a new constraint requiring the phase transition completes~\cite{Athron:2022mmm,Balazs:2023kuk} has a substantial impact on the allowed parameter space. This competes with the limit we obtain from GW observatories. We find that, in some regions of parameter space, which constraint is stronger depends on the precise treatment of the gravitational wave spectrum and resolving between them requires improvements in the understanding and precision of the gravitational wave spectra. Therefore we conclude that existing data is already sensitive to new physics from well motivated GUTs, but to fully realise the impact of this data, improvements in the theoretical predictions are necessary. Finally we also show that future runs of LIGO/VIRGO and future gravitational wave experiments will substantially extend the parameter space that can be probed in these models.

In this letter we begin by describing the details of the phase transition in the chosen PS model, including the specific field content that allows for a phase transition at the right frequency range for LIGO, and the methodolgy to compute the properties of the phase transition.  We then show the gravitational wave predictions in our model, highlighting those regions already excluded by current LIGO/VIRGO O3 results and those that will be probed or falsified in the O5 run of LIGO/VIRGO/KAGRA. Finally, we summarise our findings and discuss future applications. In this paper we focus on maximising the visibility of the GW signals at LIGO.  For a more thorough study of the model and its prediction for future missions, we refer the reader to the follow-up paper~\cite{InPrep}.

\paragraph*{Strong phase transition of the Pati-Salam model.}

PS models enhance the gauge group of the Standard Model (SM) to its PS supergroup $\mathcal{G}_{\rm PS}=SU(4)_c\times SU(2)_L \times SU(2)_R$~\cite{Pati:1974yy}. We assume that $\mathcal{G}_{\rm PS}$ spontaneously breaks into the LRSM $\mathcal{G}_{\rm LR}=SU(3)_c \times  SU(2)_L \times SU(2)_R \times U(1)_{B-L}$~\cite{Mohapatra:1974gc, Mohapatra:1974hk, Senjanovic:1975rk}, which in turn breaks down to the SM gauge group $\mathcal{G}_{\rm SM}$.  To motivate gauge coupling unification we also embed the model into an $SO(10)$ GUT~\cite{Fritzsch:1974nn} resulting in the symmetry breaking chain $SO(10)\rightarrow\mathcal{G}_{\rm PS}\rightarrow\mathcal{G}_{\rm LR}\rightarrow\mathcal{G}_{\rm SM}$.  The presence of $SU(4)$ color in $\mathcal{G}_{\rm PS}$ allows quarks and leptons to be unified into the same representation.  Within each generation, the fermions of the SM are grouped into either $\Psi_L=\{\mathbf{4,2,1}\}$ or $\Psi_R=\{\mathbf{\overline{4},1,2^*}\}$ based on chirality.  This grouping requires the existence of a right handed neutrino which facilitates the seesaw mechanism~\cite{Mohapatra:1979ia,Schechter:1980gr}.

To achieve the proposed symmetry breaking chain the minimal set of scalar fields at the PS scale is $\Phi=\{\mathbf{1,2,2}\}$, $\Delta_R=\{\mathbf{\overline{10},1,3}\}$, and $\Xi=\{\mathbf{15,1,1}\}$, which are responsible for the breaking of $\mathcal{G}_{\rm SM}$, $\mathcal{G}_{\rm LR}$, and $\mathcal{G}_{\rm PS}$ respectively.  Note that since $\Xi$ is in the adjoint representation of $SU(4)$ we can take it to be real, while the rest of the scalars are complex.  In addition to $\Delta_R$ we also include $\Delta_L=\{\mathbf{\overline{10},3,1}\}$ to facilitate the generation of neutrino masses via type II seesaw.  Lastly we include $\Omega_R=\{\mathbf{15,1,3}\}$ to explicitly break D-parity.  This allows the $SU(2)$ gauge couplings, $g_L$ and $g_R$, to run with different slopes, making it easier to achieve unification with low PS scales. Following symmetry breaking of $\mathcal{G}_{\rm PS}$ the remaining light fields make up the minimal LRSM.

\begin{figure}[htbp]
    \centering
    \includegraphics[width=\linewidth]{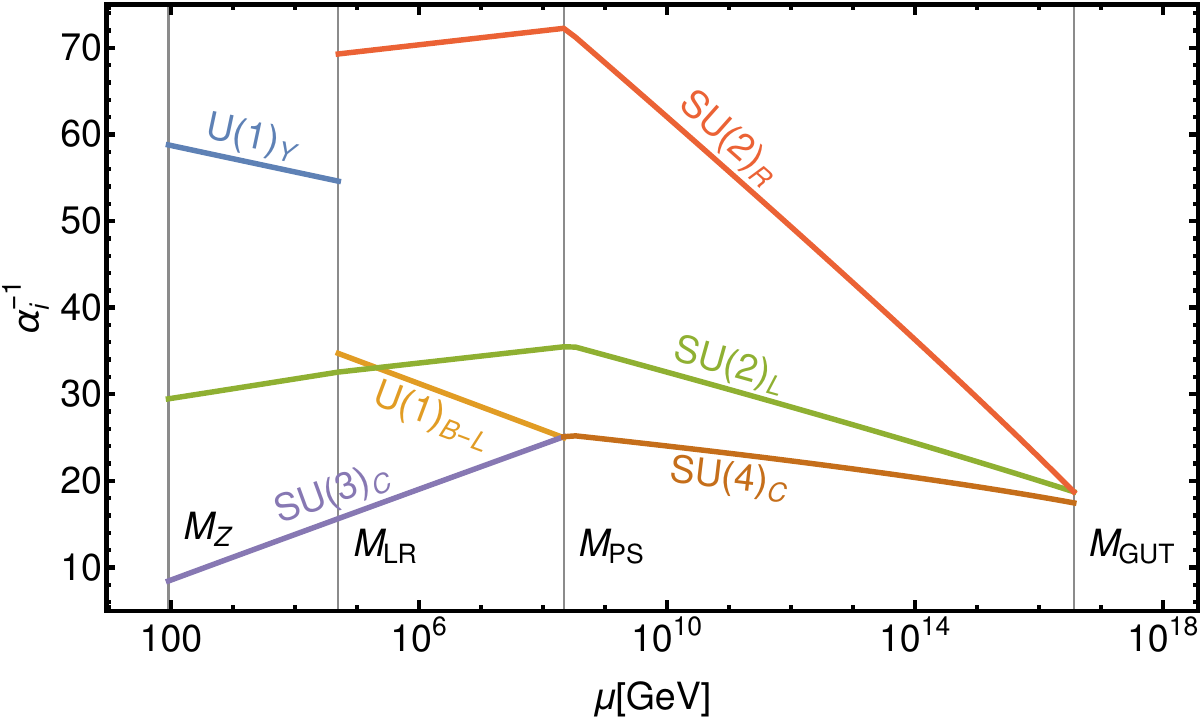}
    \caption{\emph{Gauge coupling running with renormalisation scale $\mu$ in the Pati-Salam model.}}
    \label{fig:running}
\end{figure}

To constrain the gauge coupling at the PS scale we use RGEs to run the gauge couplings from their values at the mass scale of the $Z$ boson $M_Z$, to unification at the GUT scale $M_{\rm GUT}$.  For this purpose we use {\tt PyR@te 3}~\cite{Sartore:2020gou} to compute the $\beta$-functions for the gauge couplings at two-loop, and the Yukawa couplings at one-loop, including threshold corrections~\cite{Weinberg:1980wa,Hall:1980kf,Schwichtenberg:2018cka,Meloni:2019jcf} at both the left-right ($M_{\rm LR}$) and PS ($M_{\rm PS}$) scales (see Fig.~\ref{fig:running}).
We find that couplings of order $\mathcal{O}(0.1)$ are optimal for maximising the visibility of GWs and for this choice the threshold corrections end up being quite small.
While we remain agnostic to the details of the $SO(10)$ completion, GUT scale threshold corrections could significantly relax the constraint of exact unification.  To this end we allow for partial unification, optimistically favouring situations that result in the lowest PS scale, since these are best for detectable GW signals.

In addition to constraining the model using RGEs, we also employ a few experimental limits.  At $M_{\rm LR}$, ATLAS and CMS searches for $SU(2)_R$ gauge bosons place a lower limit on the mass of $W_R$ of around $5\text{ TeV}$~\cite{ATLAS:2019isd,CMS:2021dzb}.  A similar bound could be placed on the $Z_R$ mass, but $W_R$ tends to be the lightest and hence most constrained.  In principle we could also constrain the model from GUT mediated proton decay, but the LHC constraints already constrain $M_{\rm LR}$ such that $M_{\rm GUT}$ is high enough to avoid any limits~\cite{Super-Kamiokande:2016exg}.

The PS phase transition is triggered when the fifteenth component of $\Xi$, which we denote as $\phi=\Xi_{15}$, acquires a vacuum expectation value (VEV), breaking $SU(4)$.  While in principle there are other fields present that could also acquire a VEV during the transition (i.e. a multistep transition) we assume that this is not the case.  As such the effective potential depends only on $\phi$.  We use the one-loop, daisy-resummed, temperature dependent effective potential 
\begin{equation}
    V_{\rm eff}(\phi,T)=V_{\rm tree}(\phi)+V_{\rm CW}(\phi)+V_{T}(\phi,T)+V_{\rm daisy}(\phi,T),
\end{equation}
where $V_{\rm tree}=-\frac{1}{4}\mu_{\Xi}^2\phi^2+\frac{5}{48}\lambda_{\Xi}\phi^4$ is the tree level scalar potential, $V_{\rm CW}$ is the one-loop Coleman Weinberg correction~\cite{Coleman:1973jx}, $V_{T}$ is the one-loop thermal correction~\cite{Dolan:1973qd}, and $V_{\rm daisy}$ resums higher order daisy diagrams to improve the perturbativity when $\phi^2 \ll T^2$ following the Arnold-Espinosa approach~\cite{Arnold:1992rz}.  When the mass parameter $\mu_{\Xi}^2$ is positive the symmetry is broken at tree level.  We have also replaced a particular combination of the quartic self-couplings of $\Xi$ with $\lambda_\Xi$, reducing the parameter space and ensuring the potential is bounded from below.  We set the renormalisation scale to the geometric mean of the masses of all heavy particles at the PS scale~\footnote{Uncertainties in the choice of renormalisation scale can have significant impact on the predicited gravitational wave spectrum. A thorough analysis of these uncertainties can be found in \cite{Croon:2020cgk}.}.

Since the potential only depends on one field, tracing the phase history is a relatively simple task.  We find the critical temperature directly using minimisation routines within {\tt Mathematica}.  For the nucleation rate, which depends on the $O(3)$ symmetric bounce action~\cite{Linde:1977mm,Linde:1980tt}, we pass an interpolation of our potential to {\tt BubbleProfiler}~\cite{Athron:2019nbd}.  We sample the bounce action $S_3(T)$ in the relevant temperature range and then use an interpolation to calculate the nucleation rate
\begin{equation}
	\label{eq:rate}
	\Gamma(T)\simeq T^4\left(\frac{S_3(T)}{2\pi T}\right)^{\frac{3}{2}}e^{-S_3(T)/T}.
\end{equation}

We calculate the GW signal, $\Omega_{\rm GW}$, from expressions parameterised in terms of thermal parameters~\cite{Hindmarsh:2017gnf,Caprini:2019egz}.  These are $\alpha$ (essentially a measure of the energy released from the transition normalised to the radiation energy density $\rho_R$), $R_*$ (the average bubble separation), $v_w$ (the bubble wall velocity), and $T_*$ (a reference temperature at which all quantities are evaluated).  We define $\alpha$ in terms of the trace anomaly given in e.g.~\cite{Hindmarsh:2019phv}, and, assuming the bag equation of state, express it as
\begin{equation}
    \alpha=\frac{\Delta V-\frac{1}{4}T\frac{\partial \Delta V}{\partial T}}{\rho_R(T)}\bigg\vert_{T=T_{\Gamma}},
\end{equation}
where $\Delta V=V(\phi_f)-V(\phi_t)$ is the energy difference between the true and false vacua.

The mean bubble separation $R_*$ is usually estimated from the nucleation rate by taylor expanding the bounce action \begin{equation}
    \frac{S_3(T)}{T}\approx \frac{S_3(T)}{T}\bigg\vert_{t=t_*} - \beta(t-t_*) + \frac12\beta_{V}^2(t-t_*)^2+...,
\end{equation}
\begin{equation}
    \beta =-\frac{d}{dt}\left(\frac{S_3(T)}{T}\right), \quad \beta_V = \sqrt{\frac{d^2}{dt^2}\left(\frac{S_3(T)}{T}\right)}.
\end{equation}
Keeping only the first order term implies an exponential nucleation rate $\Gamma\propto e^{\beta(t-t_0)}$ near the transition time $t_0$, from which a relation between $\beta$ and $R_*$ can be extracted.  However, we find that typically the GW signal peaks above the sensitivity window of any current or proposed detector, which means that to achieve detectable signals we need some degree of supercooling.  This is easily achieved in scenarios where a barrier persists at zero temperature. In this case the nucleation temperature (if it exists) is close to the temperature $T_{\Gamma}$ at which the nucleation rate is maximised  (corresponding to time $t_{\Gamma}$).  Around this temperature the assumption of an exponential nucleation rate breaks down as $\beta$ vanishes.  Instead we go to the next order, giving a Gaussian nucleation rate to calculate an average bubble separation
\begin{equation}
    \Gamma\propto e^{-\beta_V^2(t-t_{\Gamma})^2/2},\quad R_*\simeq\left(\frac{\beta_V}{\sqrt{2\pi} \Gamma(T_{\Gamma})}\right)^{1/3},
\end{equation}
and taking $T_{\Gamma}$ as the reference temperature $T_*$ for GW production~\cite{Megevand:2016lpr,Cutting:2018tjt,Ellis:2018mja}. 

We live in a universe where $SU(4)$ is broken and, in cases where the phase transition becomes strongly supercooled, there is a risk of the transition not completing.  To ensure that it does, we impose an approximate, analytic lower bound on the bubble wall velocity~\cite{Athron:2022mmm}.  The transition will complete so long as~\footnote{Note that in~\cite{Athron:2022mmm} the inequality is presented as an upper bound on wall velocities that do not complete.}
\begin{equation}
\label{eq:completion}
    v_w>\frac{c_f}{N^{\frac{1}{3}}(0)}\left[\sqrt{\left(\tfrac{T_{\Gamma}}{T_{\rm eq}}\right)^4+1}~{}_2F_1\left(\tfrac{1}{4},\tfrac{1}{2};\tfrac{5}{4};-\left(\tfrac{T_{\Gamma}}{T_{\rm eq}}\right)^4\right)\right]^{-1}
\end{equation}
where $N(0)$ is the total number of bubbles nucleated over the course of the transition, $T_{\rm eq}$ is the radiation-vacuum equality temperature, ${}_2F_1$ is a hypergeometric function and $c_f=\left(-3\log(f_f)/4\pi\right)^{1/3}$ is a constant related to the fraction of the universe remaining in the false vacuum, taken to be $f_f=0.01$ for completion.  We assume the bubble wall velocity approaches the speed of light, $v_w\simeq1$ as is typical of strongly supercooled phase transitions~\cite{Bodeker:2017cim}. Hence when Eq.~\ref{eq:completion} produces a lower bound greater than one, the phase transition cannot complete. 

\paragraph*{Gravitational waves and LIGO sensitivity.}

The GWs produced from the violent dynamics of first-order phase transitions can occur due to the collision of bubbles, sound waves or turbulence of the surrounding plasma. Bubble collisions are only relevant in the case of runaway bubbles~\cite{Caprini:2019egz}, and thus we neglect their contribution. We therefore calculate the GW signal $\Omega_{\rm GW}$, from sound waves and turbulence, using parametric expressions as functions of the thermal parameters $\alpha$, $R_*$, $v_w$ and $T_N$~\cite{Hindmarsh:2017gnf,Caprini:2019egz}.  In addition to the thermal parameters, we use fits to numerical simulations for the efficiency $\kappa_{\rm sw}$ at which energy liberated by the transition is converted to bulk motion in the plasma~\cite{Espinosa:2010hh}.  For the energy budget of the turbulent component, we take the conservative assumption of $\kappa_{\rm turb}\sim 0.05\kappa_{\rm sw}$.  Furthermore, we include a suppression factor $\Upsilon=1-1/\sqrt{1+2H_*\tau_{\rm sw}}$, which accounts for the finite lifetime of the sound wave source~\cite{Guo:2020grp}.  There is a large uncertainty in $\Upsilon$ coming from approximating the characteristic timescale of shock formation as $\tau_{\rm sh}\sim R_*/\overline{U}_{\parallel}$, where $\overline{U}_{\parallel}$ is the parallel component of the fluid four-velocity, which we have to further approximate as the total fluid four-velocity $\overline{U}$~\cite{Athron:2023xlk}.  We also account for reheating, using $T_{\rm rh}\simeq T_{\Gamma}[1+\alpha(T_{\Gamma})]^{1/4}$ in the redshifting of the peak frequency~\cite{Ellis:2018mja}.

Current constraints from LIGO/VIRGO (LV) come from searches in the actual O3 run data.  These look for a broken power law signal, which can be matched onto our model, combined with a background from compact binary coalesences.  No evidence has been found for either source in the data, placing limits on the strength of the GW signal.  For GWs sourced from sound waves, the relevant constraint is $\Omega_{\rm GW}(25\text{Hz})<5.7\times10^{-9}$~\cite{Romero:2021kby,Badger:2022nwo}.

To assess the detectabilty of the signal in future detectors we compute the signal to noise ratio (SNR).  To isolate the signal of a stochastic gravitational wave background from uncorrelated noise, we use cross-correlated signals in a network of detectors.  The SNR, $\rho$, in a detector network is
\begin{equation}
 \rho=\sqrt{2T}\left(\int_{f_{min}}^{f_{max}}df\sum_{I=1}^M\sum_{J>I}^M\frac{\Gamma_{IJ}(f)S^2_h(f)}{P_{nI}(f)P_{nJ}(f)}\right)^{1/2}.
\end{equation}
Here $T$ is the duration of simultaneous observation, $\Gamma_{IJ}$ is the overlap reduction function of detectors $I$ and $J$, $S_h(f)=3H_0^2\Omega_{\rm GW}(f)/(2\pi^2 f^3)$ is the GW power spectral density, computed from $\Omega_{\rm GW}$ in our model, and $P_{nI}(f)$ is the power spectral density in detctor $I$ due to noise. The sum runs over all independent pairs of detectors in the network and the integral runs over the frequency window in which the detectors are sensitive~\cite{Thrane:2013oya}.  We consider the LIGO/VIRGO/KAGRA (LVK) network operating at their O5 design sensitivities~\cite{KAGRA:2013rdx} and a proposed Einstein Telescope/Cosmic Explorer (ETCE) network~\cite{Hild:2010id,Srivastava:2022slt} both observing for one year.  Conservatively, we take the detectability threshold to be $\rho>10$~\cite{KAGRA:2013rdx}.

\begin{figure}[htbp]
    \centering
    \includegraphics[width=\linewidth]{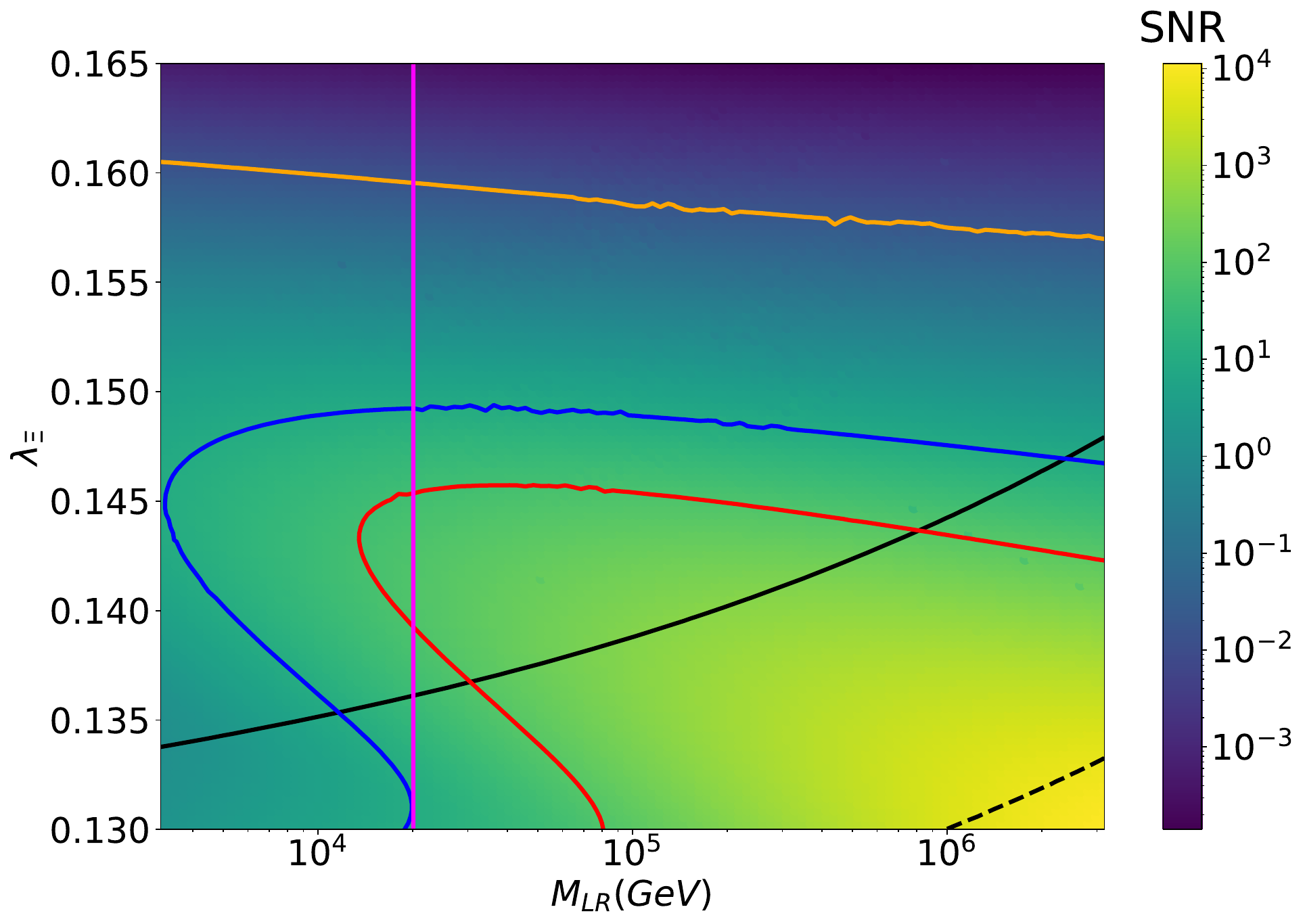}
    \caption{\emph{SNR calculated in the LIGO/VIRGO/KAGRA network operating at design sensitivity (i.e. O5 run), shown by the color gradient, with the blue contour denoting the  threshold for detection (i.e. SNR$> 10$).  The red contour is the current exclusion limit, from the LIGO/VIRGO O3 observing run, on a background produced by sound waves.  The orange contour denotes the threshold for detection for the Einstein Telescope/Cosmic Explorer.  The black contour shows a lower limit requiring that the phase transition completes.  The dashed black line is a combined constraint on $N_{\rm eff}$. The magenta line denotes a lower limit on $M_{\rm LR}$ from collider searches for heavy gauge bosons.}}
    \label{fig:results}
\end{figure}

Figure~\ref{fig:results} shows the current and future constraints from gravitational wave observatories on the parameter space of our model, in the $M_{\rm LR}$--$\lambda_{\Xi}$ plane. Excitingly, the constraints from current LV data, shown by the region enclosed below the red curve, can already exclude significant parts of parameter space. The future LVK network (O5) will have the sensitivity to either improve this constraint (as shown by the blue line), or detect the GW signal produced by the phase transition in this model. The predicted SNR for the LVK network is shown by the color gradient and highlights how the signal strength varies across the parameter space, increasing towards low values of $\lambda_\Xi$ and large values of $M_{\rm LR}$ where the transition becomes more supercooled. The future sensitivity of the ETCE network is shown by the orange line, demonstrating the significant increase in reach that is possible.

The black line represents an upper bound on the region excluded by requiring that the transition completes. This competes with the constraint from the LV network, and in fact provides a stronger constraint in the region of large $M_{\rm LR} \gtrsim 10^6$ GeV. This is because $M_{\rm LR}$ sets the VEV through the RGEs, and a larger VEV decreases the nucleation rate.  Hence a larger $\lambda_\Xi$ is required to reduce the barrier and allow the transition to complete.  The completion condition is also stronger in a small region for $M_{\rm LR} \sim 2\times 10^4$ and $\lambda_\Xi \sim 0.135$. However, very near the completion limit (for small nucleation rates) the transition may complete with very few (possibly even only one) bubbles nucleating, which goes beyond the assumptions of current simulations, highlighting the need for simulations in the strongly supercooled regime~\cite{Lewicki:2022pdb}. In spite of the uncertainties, our results demonstrate that it is crucial to include the completion criterion in analyses of supercooled transitions, as it becomes the strongest constraining factor for large portions of the parameter space, competing with and overtaking the constraints from GW observatories.

In addition to the current and predicted signals, Figure~\ref{fig:results} also shows (dashed black line) the region excluded by the combined constraint on $\Omega_{\rm GW}$ arising from strong limits on the effective number of relativistic neutrino species $N_{\rm eff}$ from the cosmic microwave background, baryon acoustic oscillations, and the deuterium abundance from big bang nucleosynthesis~\cite{Pagano:2015hma}.  Evidently, these cosmological constraints are much weaker than the requirement that the transition completes.  However, they lie in a region of the parameter space that does not complete (and hence would not produce a GW signal) so they are shown just for illustrative purposes.  Lastly the region to the left of the magenta line, corresponding to $M_{\rm LR}\lesssim 2\times10^5$ GeV, is excluded by LHC searches for $W_R$ bosons.

The region enclosed by the red, blue, black and magenta lines, roughly within $0.145\lesssim\lambda_{\Xi}\lesssim0.15$ and $2 \times 10^4 \lesssim M_{\rm LR} \lesssim \times 10^6$ is currently the most interesting section of our parameter space. It lies beyond the current excluding reach of the LV O3 network, but within reach of the sensitivity in the O5 run, presenting the possibility of a discovery in the near future. This possibility is further enhanced with the future ETCE network, which  will be able to probe higher values of the coupling up to $\lambda_\Xi \sim 0.16$ (the orange line).  

It is worth noting that the predicted GW signal, and thus the SNR shown here, is extremely sensitive to the value of $\lambda_\Xi$.  As $\lambda_\Xi$ decreases the VEV grows and so does the barrier at zero temperature.  This serves to decrease the nucleation rate, increasing $R_*$ and therefore produces a stronger signal.  However, the nucleation rate cannot be made arbitrarily small, otherwise once vacuum domination sets in, the transition will not complete.  As it turns out, the GW signal only becomes strong enough to be detectable, in current and future observatories, quite close to the parameter space region where the transition does not complete.  Therefore we have focused on the region where the signal is detectable and places a stronger limit than completion, roughly $0.13 \lesssim \lambda_\Xi \lesssim 0.165$.  This is not the case for the other model parameter $M_{\rm LR}$, on which the dependence of the GW signal is weaker, and therefore the range shown in Figure \ref{fig:results} is chosen as $3\times10^3 \textrm{ GeV} \lesssim M_{\rm LR} \lesssim 3\times10^6$ GeV, in order to focus on the region where gravitational wave observations will beat limits set by the LHC on the lower end and completion on the high end.

Figure \ref{fig:results} shows clearly the complementarity between gravitational wave observatories and collider searches, as the most interesting region lies close to the boundary where LHC searches for gauge bosons are relevant. Note that the LHC constraints are merely indicative, as they are computed approximately using $M_{W_R}\sim v_R \sim M_{\rm LR}$, where $v_R$ is the VEV of the remnant of $\Delta_R$ in the LRSM. These constraints can be weakened by shifting parameters orthogonal to those in Figure \ref{fig:results}, which will have little to no effect on the GW signal. Hence, we choose to display the worst case scenario, where the constraints from colliders are relevant in the region of interest. Further complementary probes can be extracted from the neutrino sector of the model, such as neutrino masses and lepton flavour violation, which both depend on the scale $M_{\rm LR}$. However, these also depend strongly on additional parameters and thus are less important than the collider searches shown.

\paragraph*{Conclusions and Outlook.}

Gravitational waves provide a new and unique way to probe physics at high energies that are not accessible via collider searches. In this work we have shown for the first time that LIGO/VIRGO data is already sensitive to some regions of the parameters space in well motivated models connected to grand unification.  We have also applied a new constraint, checking that the phase transition completes, which has never been applied to this class of models before.  We have shown that this constraint also has a  big impact on the allowed parameter space, competing with the constraint from LIGO/VIRGO data. We found that at higher left-right masses ($M_{\rm LR}\gtrsim 10^6$ GeV) the completion criteria is more constraining than the LIGO/VIRGO data, while for most of the mass range below this the LIGO/VIRGO data gives the strongest constraint.  Finally we demonstrated that future LIGO/VIRGO/KAGRA constraints are expected to be significantly stronger, and that the proposed Einstein Telescope/Cosmic Explorer will be able to extend this substantially, providing the strongest limit on models with left-right scales from just above the LHC limit up to very heavy scales of ${\cal O}(10^6 \text{GeV})$.  Therefore we have really entered the exciting era where GW experiments are placing constraints on well motivated grand unified theories and can provide information about high scale physics that is inaccessible to colliders.

\begin{acknowledgments}
    We thank Lachlan Morris and Eric Thrane for helpful discussions and comments.  PA acknowledges support from the National Natural Science Foundation of China (NNSFC) under grant No. 12150610460.  PA was also supported by the ARC Future fellowship FT160100274 and Discovery project DP180102209 grants in the early parts of this project. TEG is funded by the Deutsche Forschungsgemeinschaft (DFG) through the Emmy Noether Grant No. KA 4662/1-1.  MP was supported by an Australian Government Research Training Program (RTP) scholarship and a Monash Graduate Excellence scholarship (MGES).  The work of CB is also supported by the ARC Discovery Project grant DP210101636.
\end{acknowledgments}

\bibliographystyle{apsrev4-1}
\bibliography{PSGWs}

\end{document}